\documentclass[a4paper,conference]{IEEEtran}

\newcommand{\dens}{~g$\cdot$cm$^{-3}$}

\newcommand{\kernelsha}{W(\vert {\bf s}_b-{\bf s}_a\vert,h_a)}
\newcommand{\kernel}{W_{ab}(h_a)}

\usepackage{cite}      

\usepackage{graphicx}  

\usepackage{subfigure} 
\usepackage{amsmath,amssymb}   

\usepackage{fancyheadings}
\begin{document}

\title{Axisymmetric magneto-hydrodynamics with SPH}

\author{
\IEEEauthorblockN{Domingo Garc\'ia-Senz}
\IEEEauthorblockA{Departament de F\'isica,\\
Universitat Polit\`ecnica de Catalunya \\ \&
Institut d'Estudis Espacials de Catalunya \\
Barcelona, Spain\\
domingo.garcia@upc.edu}
\and
\IEEEauthorblockN{Robert Wissing}
\IEEEauthorblockA{Institute of Theoretical Astrophysics.\\
University of Oslo\\
Postboks 1029, 0315 Oslo (Norway)\\
r.b.n.h.wissing@astro.uio.no}
\and
\IEEEauthorblockN{Rub\'en M. Cabez\'on}
\IEEEauthorblockA{sciCORE,
Universit\"at Basel\\
Basel, Switzerland\\
ruben.cabezon@unibas.ch}}


\maketitle
\begin{abstract}
Many interesting terrestrial and astrophysical scenarios involving magnetic fields can be approached in axial geometry. Even though the Lagrangian smoothed particle hydrodynamics (SPH) technique has been successfully extended to handle magneto-hydrodynamic (MHD) problems, a well-verified, axisymmetric MHD scheme based on the SPH technique does not exist. In this work, we propose and check a new axisymmetric MHD hydrodynamic code that can be applied to astrophysical and engineering problems which display an adequate geometry. We show that a hydrodynamic code built on these axisymmetric premises is able to produce similar results to standard 3D-SPHMHD codes but with much lesser computational effort. 

\end{abstract}

\section{Introduction}
In spite of the large success achieved by Cartesian SPH hydrodynamic codes, there is a scarcity of SPH calculations taking advantage of the axisymmetric approach in computational fluid dynamics. To cite a few of them: \cite{petscheck93}, \cite{brook03}, \cite{garciasenz2009}, \cite{shrey19}, \cite{sun21}. But much more dramatic is the case of axisymmetric MHD simulations with SPH (SPHMHD) because, as far as we know, there is a manifest void of published material on that topic.

Nevertheless, implementing a consistent, well verified,  axisymmetric SPHMHD code may broaden the spectra of applications of such a technique. In astrophysics, the magnetic field around stellar objects can often be described with dipole or toroid geometries, both consistent with axial geometry. Relevant examples are the study of magnetized accretion disks around pulsars and the gravitational collapse of an initially spherical cloud of magnetized gas. Resolution issues add an extra degree of difficulty when these studies are conducted in three dimensions. In some cases, the axisymmetric approach is the only plausible option to study these scenarios (see, for example, \cite{Parfrey17} regarding simulations of the pulsar wind-disk interaction with an Eulerian axisymmetric hydrodynamic code). Additionally, MHD experiments in terrestrial laboratories can be largely benefited from the joint virtues of the well-established SPHMHD technique \cite{rosswog2007,pri12,price18,wissing20} plus the inherent better resolution of the axisymmetric approach. A paradigmatic example is the Z-pinch devices which aim to focus magnetically driven strong implosions towards the symmetry axis \cite{haines2000}. Additionally, researchers can take advantage of hydrodynamic codes with axial geometry to carry out convergence studies of the resolution of their own three-dimensional hydrodynamic codes.

In this work, we develop and test, for the first time,  a novel axisymmetric magneto-hydrodynamic scheme, called Axis-SPHYNX, consistent with the SPH formulation. Our proposal extends the axisymmetric code developed by \cite{garciasenz2009} to the MHD realm by adding the magnetic-stress tensor to the axisymmetric SPH equations. Furthermore, the induction and dissipative equations are consistently written in such geometry. We focus on the basic mathematical formulation of ideal MHD, so that explicit currents terms do not appear in the governing equations. 
We show that, given an axial symmetry, our MHD code is able to produce results similar to those obtained in 3D with SPHMHD codes, but with much lesser computational effort. The numerical scheme has been verified with a number of standard tests in ideal MHD, encompassing explosions/implosions, hydrodynamic instabilities, and more complex problems involving self-gravity.


\section{SPH equations of axisymmetric ideal MHD}
\label{sec-MHHForm}

  \subsection{Integral approach to estimating gradients}

Gradients and derivatives are calculated with the Integral Approach (IA) \cite{garciasenz2012} and adapted to the specificity of axial geometry. The IA approach leads to an Integral SPH scheme (ISPH), which was shown to enhance the accuracy in estimating gradients  \cite{cabezon2012,ros15,cabezon17}.  
Additionally, the ISPH formalism naturally incorporates corrective terms which are helpful in removing the so-called magnetic tensile instability. In the IA, the gradient of any scalar function $f$ carried away by particle $a$ in the axisymmetric plane defined by coordinates ${\mathbf s}(r,z)$, with $r=\sqrt{x^2+y^2}$ is, 

\begin{equation}
\left[
\begin{array}{c}
\partial f/\partial x^1\\
\partial f/\partial x^2\\

\end{array}
\right]_a
=
\left[
\begin{array}{ccc}
\tau^{11} & \tau^{12}  \\
   \tau^{21}&\tau^{22}
   
\end{array}
\right]^{-1}
\left[
\begin{array}{c}
I^1\\
I^2\\
\end{array}
\right]\,,
\label{matrix}
\end{equation}

\noindent where, from now on we use the notation $x^1\equiv r; x^2\equiv z; x^3\equiv \varphi$~(with $\varphi$~being the azimuth angle) indistinctly. Coordinate indexes $i,j,k$ are notated superscripts to make them compatible to the standard notation of the magnetic-stress tensor. Coefficients $\tau^{ij}$ $(i,j=1,2)$, and $I^i$ in Eq.~(\ref{matrix}) are,

\begin{equation}
\tau^{ij}_{a}=\sum_b \frac{m_b}{\eta_b}(x^i_{b}-x^i_{a})(x^j_{b}-x^j_{a})W_{ab}(\vert {\bf s}_b-{\bf s}_a\vert,h_a)\,,
\label{tauijsph}
\end{equation}

\begin{equation}
\begin{split}
    I({\mathbf r}_a)=&\sum_b\frac{m_b}{\eta_b} f({\mathbf r}_b)({\mathbf s}_b-{\mathbf s}_a)~W_{ab}(\vert {\bf s}_b-{\bf s}_a\vert,h_a)\\
    &- f({\mathbf r}_a)\sum_b \frac{m_b}{\eta_b} ({\mathbf s}_b-{\mathbf s}_a) W_{ab}(\vert {\bf s}_b-{\bf s}_a\vert,h_a)\,,
    \end{split}
    \label{iadfull}
\end{equation}

\noindent where $\eta_b$ is the surface density of particle $b$~ and $W_{ab}$ is the kernel function. The anti-symmetric properties of the kernel gradient, guarantee that the second term in the RHS of Eq.~(\ref{iadfull}) is close to zero. Therefore, it is neglected. That assumption gives rise to the standard ISPH scheme by \cite{garciasenz2012}. An exception to that procedure, which is connected with the magnetic tensile-instability problem, is discussed in Sect.~\ref{subsec:tensile}. 

From now on, $\kernel\equiv\kernelsha$~with $\vert {\bf s}_b-{\bf s}_a\vert=\sqrt{(r_b-r_a)^2+(z_b-z_a)^2}$~for the sake of clarity. According to \cite{cabezon2012}, the IA is related to the gradient of the kernel as,

\begin{equation}
\frac{\partial W_{ab}(h_a)}{\partial x^i_{a}}\Leftrightarrow {\mathcal A}^i_{ab}(h_a)\,; i=1,2\,,
\label{gradk}
\end{equation} 

\noindent with, 

\begin{equation}
    \mathcal A^i_{ab}(h_{a,b})=\sum^2_{j=1} c^{ij}_{a}(h_a)(x^j_{b}-x^j_{a})W_{ab}(h_{a,b})
    \,,
\end{equation} 

\noindent being $c^{ij}_{a}$~the coefficients of the inverse matrix in Eq.~(\ref{matrix}). 
We stress that although the main Axis-SPH equations are henceforth written within the ISPH formalism, translating them to the standard SPH scheme with Eq.~(\ref{gradk}) is straightforward.

\subsection{The axisymmetric SPHMHD equations} 

\vspace{0.1cm}

Adapting the axisymmetric ISPH equations to MHD is not too complicated. Volumetric ($\rho$) and surface $\eta$  densities are connected with $\eta=2\pi r \rho$. Pressure terms in the momentum equation are substituted by the magnetic stress tensor \cite{pri12}, 

\begin{equation}
    S_a^{ij}=-\left(P_a+\frac{1}{2\mu_0}B_a^2\right)\delta_{ij}+\frac{1}{\mu_0}\left(B_a^iB_a^j\right)
    \label{mhdstress}
\end{equation}

\noindent
where letter subscripts $(a,b)$~refer to particles and $\{i=1,3;~j=1,3\}$ are tensor components. Note that even though the scheme is basically two-dimensional, with coordinates ${\bf s}(r,z)$, a third coordinate, associated with the azimuth angle $\varphi$~is eventually necessary to describe the toroidal component of the magnetic field, $B_\varphi$, and velocity, $v_\varphi$. These momentum equations must also include the magnetic contribution to the hoop-stress terms, characteristic of the axisymmetric formulation \cite{brook03}. Following \cite{pri12}, the axisymmetric SPHMHD equations are built making use of the minimum action principle and the details of the procedure will be reported elsewhere. 

We write the axisymmetric SPHMHD scheme in the density averaged variant \cite{Wadsley17} because it better handles the tensile instability and allows a direct comparison with the tests cases described in \cite{wissing20}. Considering inverted reflective ghost particles in the negative semi-plane, $r<0$~(see below), guarantees that $\eta$ is correctly interpolated in the axis neighborhood.  

\begin{figure}
\includegraphics[width=0.48\textwidth]{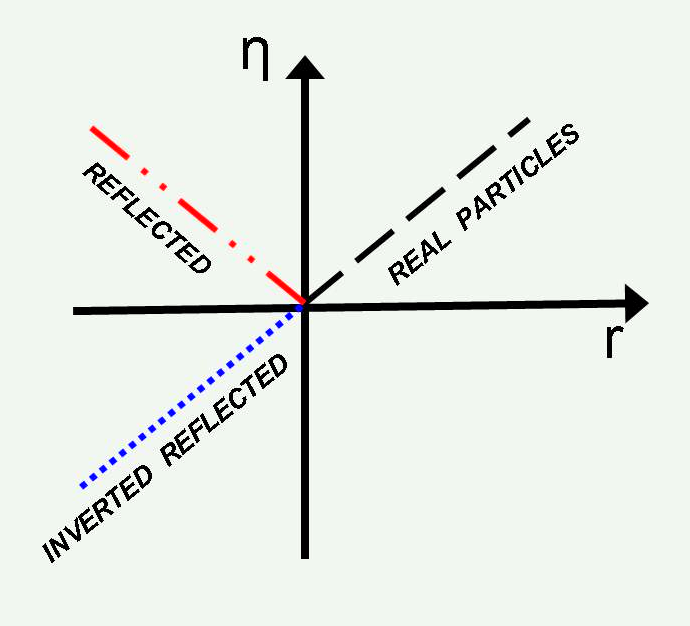}
\caption{The use of inverted-reflected ghosts particles along with the IA technique overcomes the numerical troubles when calculating the density $\eta$ and its gradient near the symmetry axis.}
\label{fig:invrefl}
\end{figure}
\medskip
\begin{itemize}
    \setlength\itemsep{0.2cm}
    \item {\sl Mass conservation}
    
    \begin{equation}
\eta_a=\sum_{b=1}^N \varepsilon_b~m_b W_{ab}(h_a)\,,
\label{densitymhd}
\end{equation}
    
 \noindent where $\varepsilon_b=\pm 1$ is a parameter which assigns a signature to the neighbor particle. Real particles have $\varepsilon_b =+1$ whereas ghost particles across the axis have $\varepsilon_b = -1$. According to Fig.~\ref{fig:invrefl}, the use of such inverted-reflected particles ensures that $\eta$ behaves linearly when $r\rightarrow 0$ and is, therefore, correctly interpolated. The signature $\varepsilon$~also affects the momentum and energy equations.
    
   \item {\sl Momentum equations}
   
   \begin{equation}
\begin{split}
a^r_a &= 2\pi\frac{\left(P_a+\frac{B_a^2}{2\mu_0}-\frac{(B_a^\varphi)^2}{\mu_0}\right)}{\eta_a}+\\
&2\pi \sum_{b=1}^N  m_b \left(\frac{S_a^{ri} \vert r_a\vert}{\eta_a\eta_b}{\mathcal A}_{ab}^{i}(h_a)+
 \varepsilon_b\frac{S_b^{ri} \vert r_b\vert}{\eta_a\eta_b} {\mathcal A}_{ab}^{i}(h_b)\right)\,.
\end{split}
\label{mhdaccel_r}
\end{equation}

\begin{equation}
a^z_a= 2\pi\sum_{b=1}^N m_b\left(\frac{S_a^{zi} \vert r_a\vert}{\eta_a\eta_b}{\mathcal A}_{ab}^{i}(h_a)+\varepsilon_b\frac{S_b^{zi} \vert r_b\vert}{\eta_a\eta_b}{\mathcal A}_{ab}^{i}(h_b)\right)\,.
\label{mhdaccel_z}
\end{equation}

\begin{equation}
\begin{split}
a^\varphi_a & = 2\pi\left(\frac{B_a^r B_a^{\varphi}}{\mu_0\eta_a}\right)+\\
&2\pi\sum_{b=1}^N  m_b \left(\frac{S_a^{\varphi i} \vert r_a\vert}{\eta_a\eta_b}{\mathcal A}_{ab}^{i}(h_a)+
 \varepsilon_b\frac{S_b^{\varphi i} \vert r_b\vert}{\eta_a\eta_b} {\mathcal A}_{ab}^{i}(h_b)\right)\,,
\end{split}
\label{mhdaccel_phi}
\end{equation}

\noindent where $\{a^r_a, a^z_a,a^\varphi_a\}$ are the acceleration components in cylindrical coordinates and repeated indexes, $i (=r,z)$ are summed. Equation~(\ref{mhdaccel_phi}) is only relevant in those applications involving both, \{$v_\varphi, B_\varphi\neq 0$\}, as is the case of scenarios combining rotation and toroidal magnetic fields. Its impact in the simulations is discussed in Sect.~\ref{subsec:collapse}.

\item {\sl Energy equation}

\begin{equation}
\frac{du_a}{dt}=-2\pi\frac{P_a}{\eta_a} v_{r_a}+
2\pi\frac{P_a \vert r_a\vert }{\eta_a}\sum_{b=1}^N \frac{m_b}{\eta_b} \left( v^i_{ab}~{{\bf\mathcal A}^i_{ab}}(h_a)\right)\,.
\label{energy}
\end{equation}

\end{itemize}

\subsection{The induction equation}
\label{subsec:induction}

The induction equation is first written similarly to \cite{pri12},

\begin{equation}
\frac{d{\bf B}}{dt} = -{\bf B}({\bf\nabla}\cdot{\bf v}) + ({\bf B}\cdot {\bf \nabla}){\bf v}\,,
\label{induction}
\end{equation}

\noindent
where the non-ideal term associated with the current density $\bf J$ has been taken out from the expression. Writing ${\bf B}({\bf\nabla}\cdot{\bf v})$ and the material derivative $({\bf B}\cdot {\bf \nabla}){\bf v}$ in cylindrical coordinates, taking $\frac{\partial}{\partial \varphi}=0$, and manipulating, we have,

\begin{equation}
 \begin{split}
 &\frac{d}{dt} 
\begin{bmatrix}
 B^r \\ 
  B^z \\
 B^\varphi 
\end{bmatrix}
= \\
    &
\begin{bmatrix}
-\left(\frac{\partial v_z}{\partial z}+\frac{v_r}{r}\right)& \frac{\partial v_r}{\partial z}& -\frac{v_\varphi}{r} \\
\frac{\partial v_z}{\partial r}& -\left(\frac{\partial v_r}{\partial r}+\frac{v_r}{r}\right) & 0\\
\frac{\partial v_\varphi}{\partial r}&\frac{\partial v_\varphi}{\partial z} &-\left(\frac{\partial v_r}{\partial r}+\frac{\partial v_z}{\partial z}\right) \\
\end{bmatrix}
\begin{bmatrix}
 B^r \\ 
  B^z \\
 B^\varphi 
\end{bmatrix}\,.
\end{split}
\label{matrixind}
\end{equation}  

Thus, the induction equation is written as a linear equation,

\begin{equation}
\frac{d{B^i_{a}}}{dt} = \sum_{j=1}^3 r^{ij} B^j_{a}\,,
\label{linearinduction}
\end{equation}

\noindent
where the coefficients $r^{ij}$ only depend on the velocity field around the particle.

\subsection{Dissipation}
\label{subsec:dissipation}

As in Cartesian SPH, the axisymmetric approach needs some amount of dissipation to handle shock waves. As usual in SPH, this is done with the artificial viscosity (AV) formulation. There are two main sources of dissipation in MHD: those from the AV and those arising from the induced currents in plasma sheets during collisions. The former is purely hydrodynamical and is the same as that implemented in SPHYNX \cite{cabezon17,gsenz22} with the third coordinate removed. Only the Balsara limiters \cite{balsara95} to AV have been included in the present version of the code. For the latter, we use the scheme described in \cite{wissing20}, 

\begin{equation}
    \left(\frac{d\mathbf B}{dt}\right)^{diss}=\xi_B \nabla^2\mathbf B\,,
    \label{Bdiss_1}
\end{equation}

\noindent with $\xi_B = \alpha_B~v_{sig,B}~h$, mimicking a magnetic resistivity parameter,  $v_{sig,B}$ being the characteristic signal velocity, and $\alpha_{B}\simeq 1$. The numerical analog of Eq.~(\ref{Bdiss_1}) 
has a Cartesian-like contribution (but with coordinates $r,z$),

\begin{equation}
    \left(\frac{d\mathbf B}{dt}\right)_a^{diss,C}=\sum_{b=1}^{n_b} \frac{m_b}{\eta_b}~ \frac{\xi_{B,a}+\xi_{B,b}}{\vert s_{ab}\vert}\mathbf B_{ab}\left(\hat s^i_{ab}\tilde A^i_{ab}\right)\,,
    \label{Bdiss_2}
\end{equation}

\noindent where ${\mathbf B}_{ab}={\mathbf B}_a-{\mathbf B}_b$, $\hat s_{ab}$ is the unit vector joining the particles $a,b$ in the axisymmetric plane and $\tilde A^i_{ab}=0.5 [A^i_{ab}(h_a)+A^i_{ab}(h_b)]$.


In cylindrical geometry there are other contributions to be added to the Cartesian part (Eq.~\ref{Bdiss_2}). The complete expression to compute each component of the magnetic dissipation is, 

\begin{equation}
\begin{split}
   \left(\frac{dB^i}{dt}\right)_a^{diss} =& \left(\frac{d B^i}{dt}\right)_a^{diss,C}+\left(\frac{\xi_B}{r}~\frac{\partial B^i}{\partial r}\right)_a \\ &- (1-\delta^{iz}) \left(\frac{\xi_B }{r^2}B^i\right)_a\,,
   \end{split}
\end{equation}

\noindent where $\delta^{iz}~ (i=r,z,\varphi)$~is the Kronecker-delta function. The contribution of such `non-Cartesian' terms in the test cases below was, however, subdominant and was neglected.\footnote{Additionally, it is not evident how to build a suitable contribution  of these terms to the energy equation,  Eq.~(\ref{Bdiss_3}).}

According to \cite{wissing20}, the magnetic dissipation contributes to the rate of change of internal energy, Eq.~(\ref{energy}) as, 
 
\begin{equation}
     \left(\frac{du}{dt}\right)_a^{diss}=-\frac{\pi r_a}{\mu_0\eta_a}\sum_{b=1}^{n_b}\frac{m_b}{\eta_b}~ \frac{\xi_{B,a}+\xi_{B,b}}{\vert s_{ab}\vert}\mathbf B_{ab}^2\left(\hat s^i_{ab}\mathcal{\tilde A}^i_{ab}\right)\,,
    \label{Bdiss_3}
\end{equation}
 
In the tests below, the adopted value of $\xi_B$ is, 
 
\begin{equation}
     \xi_B=\frac{1}{2}\alpha_B~v_{sig,B}\vert s_{ab}\vert\,.
\end{equation}
    
For the signal velocity we take the expression by \cite{price18}, 

\begin{equation}
    v_{sig,B}=\vert \mathbf v_{ab}\times \hat{\mathbf s}_{ab}\vert
    \label{Bdiss_4}\,
\end{equation}

\subsection{Removing the magnetic tensile instability}
\label{subsec:tensile}

Calculations where magnetic pressure largely exceeds the kinetic gas pressure are prone to undergo the tensile instability \cite{phi85}. Such instability concerns the nocive effect of the magnetic-stress tensor elements $B^i B^j/\mu_0$, when they become dominant. The tensile instability manifests in the artificial clumping of particles and is often the source of numerical troubles. One of the first solutions to getting rid of this instability was suggested by \cite{Morris96}, who subtracted the last term in the RHS in Eq.~(\ref{mhdstress}) from the acceleration equation, Eqs.~(\ref{mhdaccel_r},\ref{mhdaccel_z}). Other expressions of such corrective term to the acceleration can be found in \cite{borve01}, \cite{pri12}.

It is worth noting that the ISPH scheme provides a similar corrective term to that by \cite{Morris96}. The idea is to take into account the last term in the RHS of Eq.~(\ref{iadfull}) to build a suitable corrective term to the acceleration. According to \cite{garciasenz2012} such term, $f^i_{\nabla{B},a}$ is, 

\begin{equation}
 f^i_{\nabla{B},a} = - 2\sum_b m_b \frac{(\frac{B^iB^j}{\mu_0})_a}{\rho_a\rho_b}{\nabla_a^j W}_{ab}(h_a)\,, 
 \label{tensile_1}
\end{equation}

The corrective term is steered by the parameter $\beta_a=\frac{2\mu_0 P_a}{B_a^2}$, and to smooth the transition between the weak and strong field regimes interpolating function by \cite{wissing20} has been used, 

\begin{equation}
{\mathcal H}_{a}=
\left\{\begin{array}{rclcc}
+2 & \qquad \beta_a < 1 \\
2(2-\beta_a) & \qquad 1\le\beta_a \le 2\\
0 & \qquad\mathrm {Otherwise}, 
\end{array}
\right.
\label{betainterp}
\end{equation}

Equation~(\ref{tensile_1}), is easily adapted to axial geometry as,

\begin{equation}
 f^i_{\nabla{B},a} = -2\pi \mathcal{H}_a\sum_b m_b \frac{(\frac{B^iB^j}{\mu_0})_a \vert r_a\vert}{\eta_a\eta_b}{{\mathcal{A}}}^j_{ab}(h_a)\,. 
 \label{tensile_2}
\end{equation}

The magnitude $f^i_{\nabla{B},a}$ in Eq.~(\ref{tensile_2}) is added to Eqs.~(\ref{mhdaccel_r},\ref{mhdaccel_z}) to obtain the acceleration of the SPH-particle.

\subsection{Cleaning the divergence of B}
\label{sec:cleaning}

A challenge of numerical MHD is to permanently fulfill the condition $\mathbf\nabla\cdot \mathbf B = 0$. In most existing SPH codes this is achieved with divergence cleaning techniques. Here we use the hyperbolic/parabolic cleaning scheme by \cite{tricco12} which has proven very satisfactory to keep $\nabla\cdot{\mathbf B}$ at negligible levels. Adapting this cleaning scheme to the axisymmetric geometry is straightforward. According to \cite{wissing20}, the cleaning parameters are set to $f_{clean}=1$, ${\sigma}_c=1$.

\section{Tests}
\label{sec-models}

We describe the implementation and results of four tests encompassing a variety of physical phenomena such as explosions, implosion, instabilities, and gravitational collapse. On the whole, we found a good match between the axisymmetric code Axis-SPHYNX and the results obtained with the 3D-hydrodynamic code GDSPH by \cite{wissing20}. 

The equation of state (EOS) was that of an ideal gas with $\gamma=5/3$, except in the last test where a barotropic EOS was used. To build initial models with homogeneous density in cylindric geometry we spread the SPH particles with mass $m_a^0$~in a uniform squared grid and re-normalize the value of their mass according to $m_a=m^0_a~ r$. While such a simple recipe was enough for the purposes of this work, more elaborated initial models could be necessary for other applications. The magnetic permittivity was taken $\mu_0=1$.
Information regarding the chosen value of several parameters in Axis-SPHYNX is shown in Table \ref{tab:table1}.

\begin{table}
	\centering
	\caption{Default value of relevant simulation parameters with Axis-SPHYNX. Columns are:  number of neighbors $n_b$, AV coefficients, heat diffusion coefficient ($\alpha_u$) in AV, magnetic dissipation coefficient ($\alpha_B$) and cleaning parameters.}
	\begin{tabular}{lccccccr} 
		\hline
		$n_b$& $\alpha_{AV}$&$\beta_{AV}$&$\alpha_u$&$\alpha_B$&$f_{clean}$&$\sigma_{clean}$& \\
		\hline
		$60$&$1$&$2$&$0.05$&$0.5$&$1$&$1$&\\
		\hline
	\end{tabular}
	\label{tab:table1}
\end{table}

  \subsection{The magnetic Sedov test}
\label{subsec:sedov}

\begin{figure}
\includegraphics[width=0.50\textwidth]{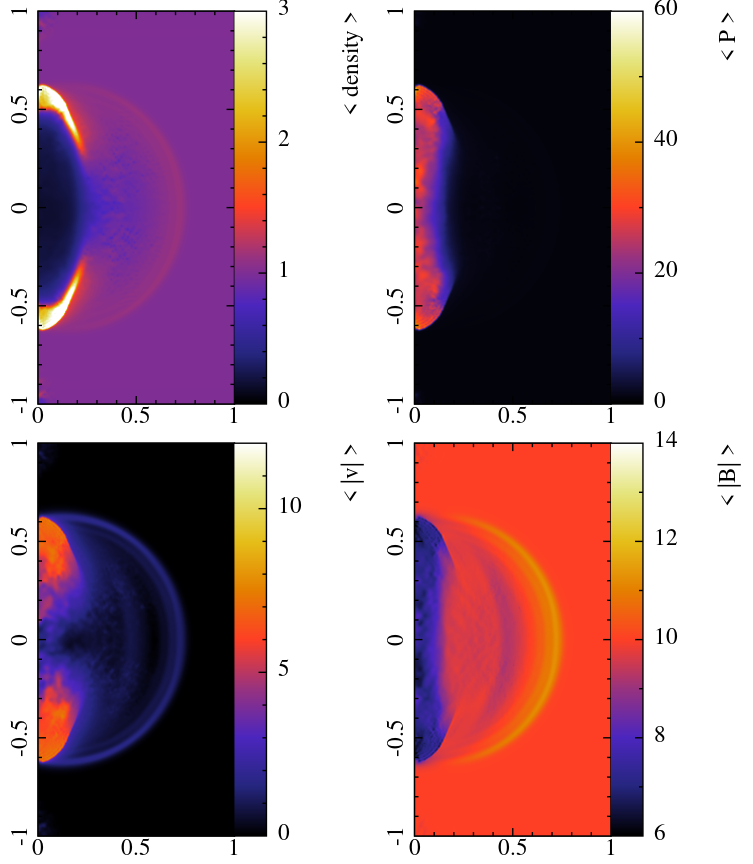}
\caption{Point-like explosion in a magnetized medium calculated with Axis-SPHYNX at time $t=0.048$}
\label{fig:sedov_1}
\end{figure}

The axisymmetric version of the MHD Sedov test is easily set by considering an initially spherically symmetric explosion amidst a uniform magnetic field $\mathbf B(r,z)= B_z \hat z$. We compare the evolution computed with Axis-SPHYNX to that obtained with GDSPH for the same initial conditions, but in 3D. To seed the explosion a Gaussian initial profile of internal energy was assumed: 

\begin{equation}
    u(s)= u_0 \exp[-(s/\delta)^2]+ u_b\,,
    \label{sedov_1}
\end{equation}

\noindent with $s=\sqrt{r^2+z^2}$~and, 

\begin{equation}
u_0= \frac{E_{tot}}{(\pi^{3/2}~\rho_0~\delta^3)}\,, 
\end{equation}

 \noindent where $E_{tot}=5$~is the total initial energy of the explosion, $\delta=0.1$, and ${\mathbf B} = 10~\hat {\bf z}$. The medium was initially homogeneous with $\rho_0=1$ and a background internal energy $u_b=1$. Periodic boundary conditions were implemented in the 3D calculation and a mix of reflective, (left and right planes), and periodic (up and down planes), in the axisymmetric approach. The number of particles was $N=362^2$ (average smoothing length, $h\simeq 8\cdot 10^{-3}$), in the Axis-SPHYNX calculation and $N=125^3~ (h\simeq 22\cdot 10^{-3}$), in the GDSPH run. 
 
\begin{figure}
\includegraphics[width=0.51\textwidth]{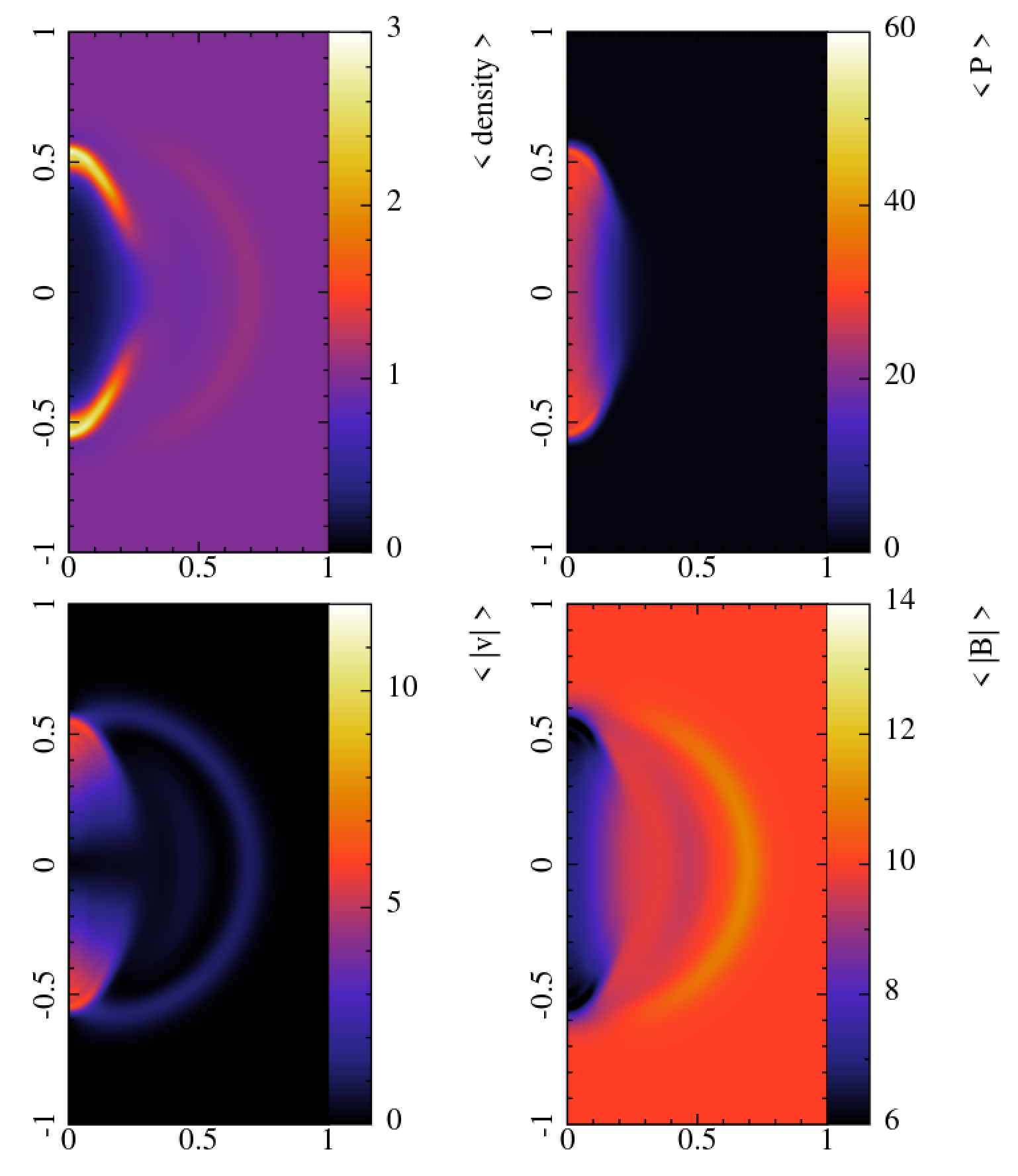}
\caption{Point-like explosion in a magnetized medium calculated with GDSPH. There are shown the same magnitudes as in Fig.~\ref{fig:sedov_1} but in a cut in the plane Z with thickness $h$.}
\label{fig:sedov_2}
\end{figure}

 The results of the calculations are summarized in Figs.~\ref{fig:sedov_1} and \ref{fig:sedov_2}. The color maps, depicting density, pressure, and modulus of velocity and magnetic field at $t=0.048$, do not reveal significant differences between the simulations performed by Axis-SPHYNX (Fig.~\ref{fig:sedov_1}), and GDSPH (Fig.~\ref{fig:sedov_2}). They also look qualitatively similar to the results published by other authors \cite{rosswog2007}. As shown in the first panel in Figs.~\ref{fig:sedov_1} and \ref{fig:sedov_2}, the shock front is slightly ahead in the axisymmetric calculation, which is due to the higher resolution in that calculation. The color maps of velocity (bottom-left panels) show extended regions with low velocity (blue regions) in both cases. Close to the symmetry axis the distribution of particles is, however, less ordered in the axisymmetric calculation, where the velocity fluctuates more. The total energy, $E$, is well conserved, $\frac{\Delta E}{E_0}\simeq 10^{-2}~\%$ at all times. The value of the estimator $\epsilon_{div}=\left<\frac{h\nabla\cdot{\mathbf B}}{\vert {\mathbf B}\vert}\right>$, measuring the averaged relative error of the constraint $\nabla\cdot{\mathbf B}=0$ was always $\epsilon_{div}\le 0.2\%$.

\subsection{Z-pinch like implosion}
\label{subsec:alfven}

Z-pinch devices were among the first to explore the feasibility of having controlled nuclear fusion in terrestrial laboratories (see \cite{haines2000,shumlak20} for a review). They have also been applied to conduct laboratory astrophysics experiments \cite{Bocchi13}. In the Z-pinch machines a strong toroidal magnetic field, $B^\varphi$ is created by a mega-amp\`ere electric current pulse moving along the Z-axis. The Lorentz force exerted by $B^\varphi$~on the plasma, which initially moves in the Z-direction, impels it towards the Z-axis, provoking the implosion. Provided that the initial conditions are axisymmetric, the compression of the plasma at the symmetry axis can be strong.  

To arrange a Z-pinch magnetic implosion in a simple numerical experiment, we consider an initially homogeneous plasma with $\rho=1, P=1$ in a cylinder with radius $R=1$, and height $Z=2$. The plasma is initially moving with $v^z= -1$. We set a toroidal magnetic field, $B^\varphi$, with maximum strength $B^\varphi_0=3$ and with a gaussian profile,  

\begin{equation}
    B^\varphi= B^\varphi_0~\exp\left[-(r-r_0)^2/\delta\right]\,,
    \label{zpinch_1}
\end{equation}

\noindent with a characteristic width $\delta=0.01$. The boundary conditions are periodic on the top and bottom of the cylinder and reflective on the lateral surfaces. As in the Sedov-test, we aim to compare the results of Axis-SPHYNX to those obtained with the three-dimensional code GDSPH, and identical initial conditions. In this test, both hydrodynamic codes have a similar initial resolution, $h\simeq 8\cdot 10^{-3}$, but with $N^{2D}=362^2$ and $N^{3D}=512^2\times 24$ particles in the axisymmetric and full 3D calculations, respectively. 

\begin{figure*}
\begin{center}
\includegraphics[width=0.8\textwidth]{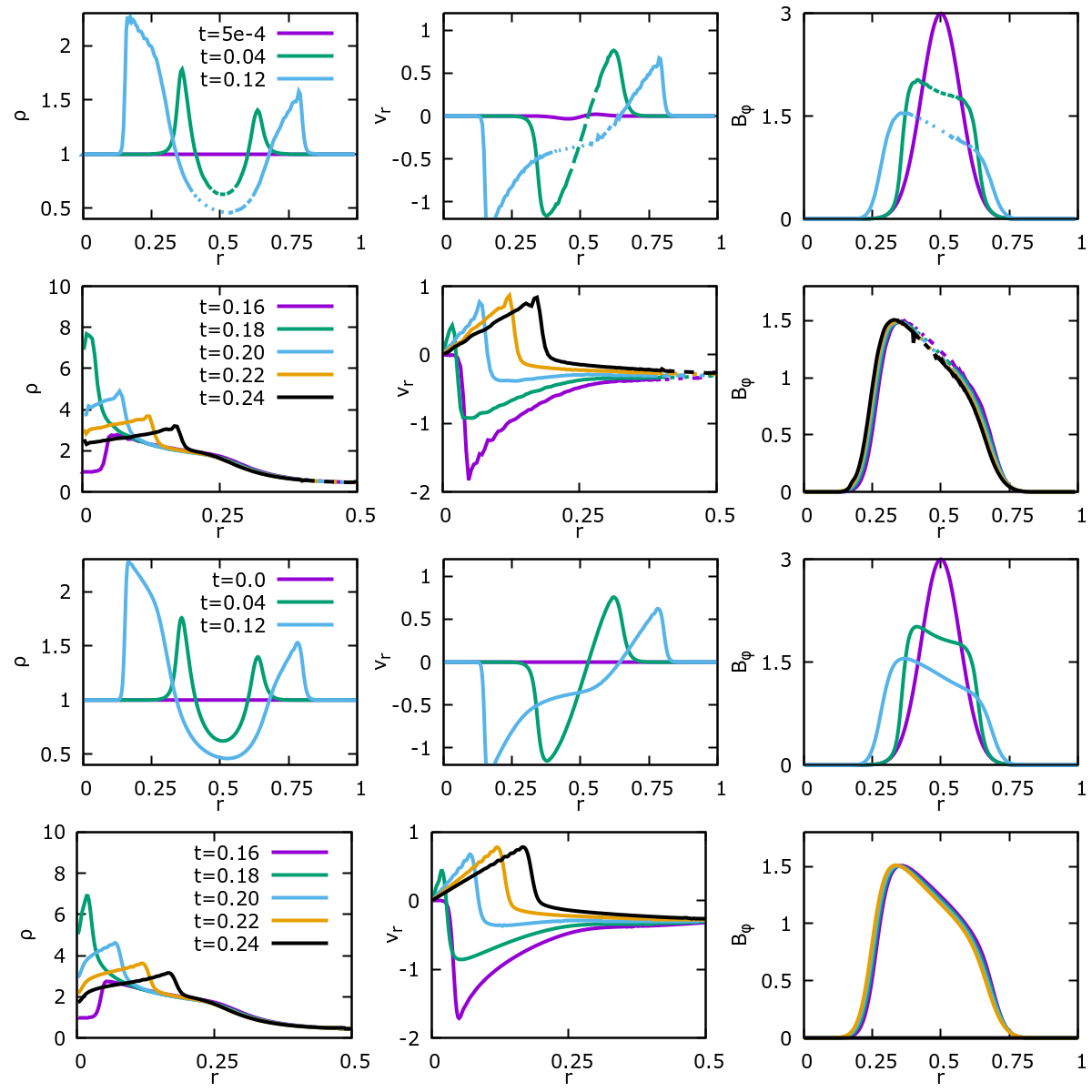}
\end{center}
\caption{Z-pinch-like implosion in a magnetized medium calculated with Axis-SPHYNX (first and second rows of panels) and with GDSPH (third and fourth rows of panels). The figure shows the shell-averaged values of density $\rho$, radial velocity $v^r$~and toroidal component, $B^\varphi$ of the magnetic field.}
\label{fig:zpinch_1}
\end{figure*}

The main results of this numerical experiment are shown in Fig.~\ref{fig:zpinch_1}. That figure depicts the profiles of $r-$averaged magnitudes, $\rho$, $v^r$, and $B^\varphi$, at different elapsed times. The first and second rows correspond to the axisymmetric calculation whereas the lower two resulted from the full three-dimensional calculation with GDSPH. As we can see, the match between the results obtained with both codes is excellent. The density peak around the point of maximum compression at $t=0.18$ is a bit larger in the axisymmetric calculation. The radial velocity profile at the supersonic shock front is sharp and well captured in both calculations. The $\left<v^r\right>$ profiles obtained with Axis-SPHYNX are a bit noisier than those with GDSPH, probably due to the lower number of neighbors, $n_b\simeq 60$, used to carry out the interpolations. The toroidal component of the magnetic field evolves very similarly in both calculations.  
The total energy was preserved up to $\frac{\Delta E}{E_0}\le 0.4\%$ and the constraint $\nabla\cdot{\mathbf B}=0$ was fulfilled to machine precision. 

\subsection{Magnetic Kelvin-Helmholtz instability}
\label{subsec:KH}

The growth of the Kelvin-Helmholtz instability across the contact layer between fluids with different densities is a challenging test for hydrodynamic codes. Resolution issues limit the growth rate of the instability during the initial linear stage which, later on, hinders the development of small wave lengths in the non-linear phase \cite{mcn12}. Modern SPH codes are able to cope with the KH instability but only if the number of particles is high, several millions as a minimum (in 3D), and the density contrast is usually not very large. Adding a magnetic field to the plasma turns this test into an interesting, albeit more complex, MHD problem. 
SPH 3D simulations of the growth of the KH instability in a weakly magnetized media have been reported by \cite{hopkins16,wissing20}, among others. The main effect of the magnetic field is to uncoil and stretch the vortex during the non-linear stage so that the instability looks rather different from that of non-magnetized systems. The axisymmetric realization of these 3D-MHD experiments consists of two interacting fluids moving along two concentric cylindrical pipes, but in opposite directions. A uniform magnetic field, $B^z$, pointing along the axis of the pipe, is added so that it interacts with the radial component of the velocity $v^r$ in the unstable layer, via the Lorentz force.

We consider a cylinder with radius $R=1$ and length $L=2$. A fluid with density $\rho_{in}=2$ moving with $v^z=+0.5$ fills the inner half, $r\le R/2$, of the cylinder. The outer part of the cylinder is filled with a lighter fluid, $\rho_{out}=1$, moving with $v^z=-0.5$. Both fluids share the same pressure, $P=2.5$ and are immersed in a magnetic field $B^z=0.1$. The fluid interface is altered by adding a small radial perturbation to $v^r$,       

\begin{figure*}
\includegraphics[width=1\textwidth]{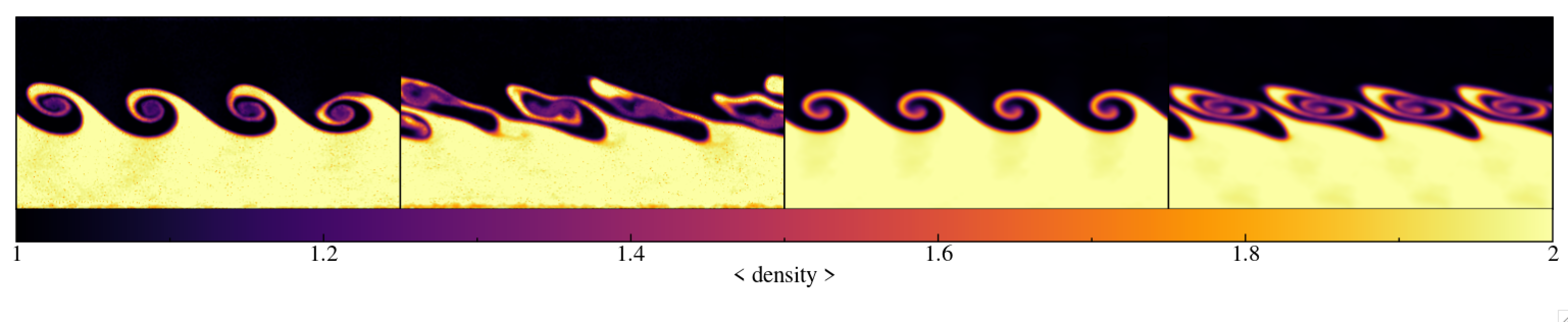}
\caption{Particles distribution at two elapsed times, $t=1.5$~and $t=2.8$, for models calculated with Axis-SPHYNX (first two panels) and  calculated with GDSPH (XZ slices in the two rightmost panels).}
\label{fig:kh_1}
\end{figure*}

\begin{equation}
    v^r=\Delta v^r\exp\left(-\frac{\vert r-0.5\vert}{0.1}\right)\sin\left(4\pi z\right)
    \label{kh_2}
\end{equation}

\noindent with $\Delta v^r =0.05$. The number of particles in the Axis-SPHYNX and GDSPH runs was $N=422^2$ and $N=256^3$, respectively. Figure~\ref{fig:kh_1} depicts the density color-map at two times, $t=1.5$ and $t=2.8$, being the former representative of the hydrodynamic stage and the latter of the time when MHD effects take over (the characteristic growth time-scale is $\tau_{KH}\simeq 1.06$). The match between both codes at $t=1.5$ is good, with qualitatively similar development of the structures and sub-structures. In the long run, the magnetic field manages to distort the morphology of the billows and vortexes. At $t=2.8$ the morphology of the billows (second and fourth snapshots in Fig.~\ref{fig:kh_1}) is qualitatively similar. In both cases, the MHD effects stretch the vortex, but the flow loses the symmetry faster in the axisymmetric calculation. Energy is conserved to ${\Delta E}/{E_0}\le 0.1\%$ whereas the divergence constraint is satisfied up to $\epsilon_{div}\le 2\%$.

\subsection{Collapse of a rotating-magnetized cloud}
\label{subsec:collapse}

The collapse of a rotating and magnetized dense cloud of gas embedded in a more dilute medium has become a standard test to verify MHD hydrodynamic codes \cite{hopkins16}. This test involves many physical ingredients of astrophysical interest such as gravity, rotation, and magnetic fields. Because the collapse of the cloud basically proceeds with axial geometry (except in those cases where there is fragmentation) this scenario can be approached with axisymmetric MHD codes. 

A cloud with mass $M=1~M_{\odot}$ and density $\rho_0= 4.8\cdot 10^{-18}$\dens~rotates around the Z-axis with $\omega_0= 4.24\cdot 10^{-13}$~s$^{-1}$. The cloud is surrounded by a  rarefied medium, with a radius ten times larger than that of the cloud and density $\rho_M=\rho_0/300$. The whole system is inside a magnetic field $\mathbf B = \frac{610}{\mu} \hat z~\mu G$ aligned with the rotation axis of the cloud. Three-dimensional simulations of the collapse, with a barotropic EOS, have shown that the implosion of the cloud would produce a narrow jet only if the parameter $\mu$ is neither too large ($\mu\le 75$), nor too small ($\mu\ge 2$) \cite{hopkins16, wissing20}. 

This test is extremely challenging to an axisymmetric SPH code because the collapse is fierce and impels the particles towards the singularity axis. The central density increases five orders of magnitude and the Courant criterion enforces the time-step to be really small. We have carried out three simulations of this scenario with $\mu=\infty, \mu=20, \mu=10$, from the initially spherically symmetric conditions until the formation of the disk and beginning of the jet launching at $t\simeq 1.1\cdot 10^{12}$~s, which is close to the characteristic free-fall time of the cloud. 

The gravity force (${\mathbf g}$) is calculated using the scheme described in \cite{garciasenz2009} and is added to the acceleration. For this problem it is better to seek for the angular velocity, $\omega ({\mathbf s}, t)$ rather than for $v^\varphi$. The momentum equations, Eqs.~(\ref{mhdaccel_r},\ref{mhdaccel_z},\ref{mhdaccel_phi}), become, 

\begin{equation}
    \frac{dv^r_a}{dt}= a^r_a + g^r_a + \omega_a^2 r_a\,.
    \label{cloud_r}
\end{equation}

\begin{equation}
    \frac{dv^z_a}{dt}= a^z_a + g^z_a\,.
    \label{cloud_z}
\end{equation}

\begin{equation}
    \frac{d\omega_a}{dt}=\frac{1}{r_a} a^\varphi_a - 2\frac{v^r_a~\omega_a}{ r_a}\,.
    \label{cloud_phi}
\end{equation}

\begin{figure}
\includegraphics[width=0.50\textwidth]{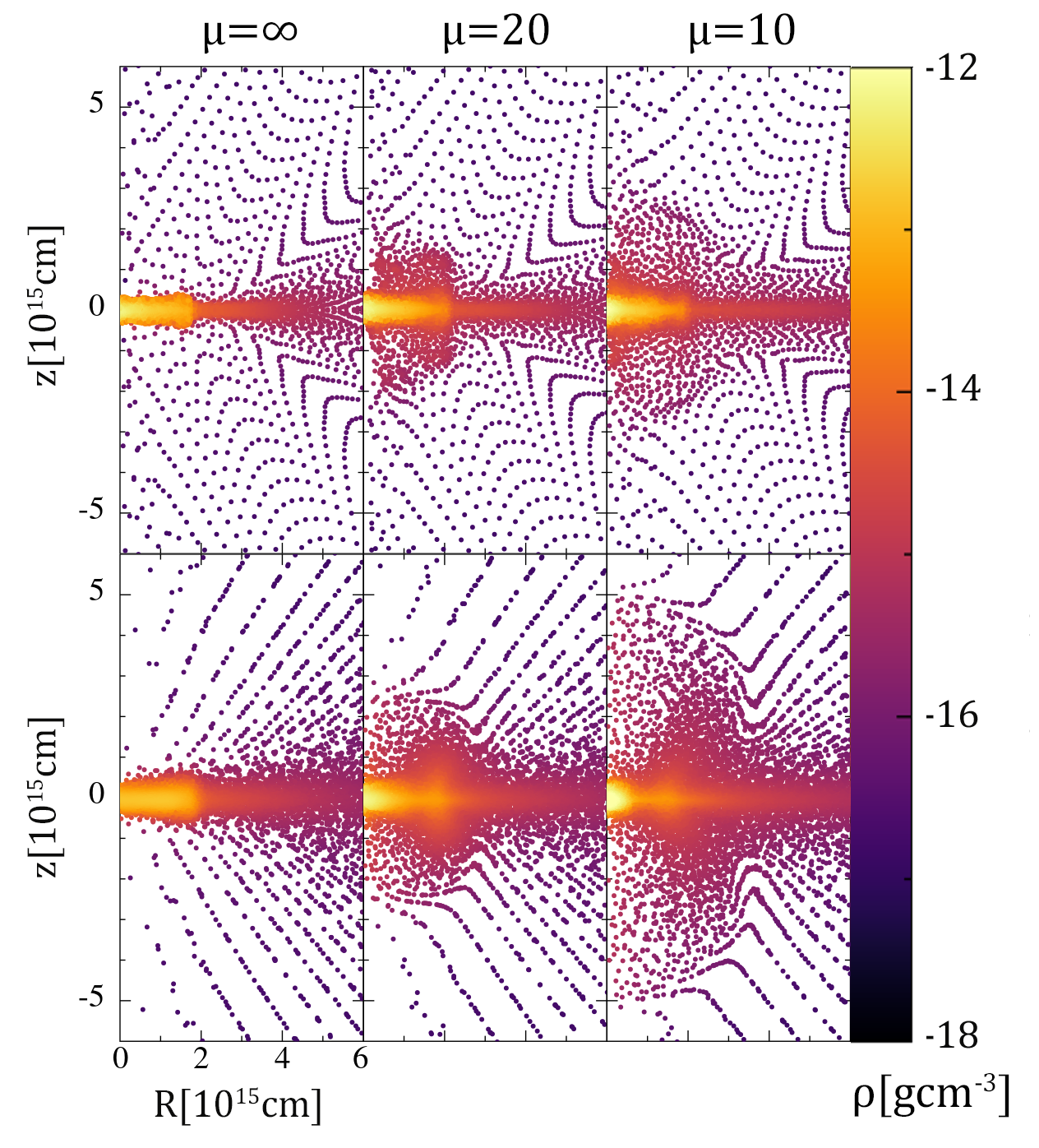}
\caption{Density color-maps of the core  of the collapsing cloud  at $t=1.1\cdot 10^{12}$~s. Upper panels show the results with Axis-SPHYNX for three values of the magnetic field, $B^z=610/\mu$. The same is shown in the lower panels, but calculated with the code GDSPH.}
\label{fig:cloud1}
\end{figure}

The case with $a^\varphi_a=0$ is equivalent to the conservation of angular momentum. Figure~\ref{fig:cloud1} shows the density color-map of the innermost region of the cloud at $t=1.1\cdot 10^{12}$~s,  when the jets, if any, are born. The upper panels depict the calculation with Axis-SPHYNX\footnote{Because, for now,  gravity is  calculated computing direct particle-particle interactions \cite{garciasenz2009}  the number of particles used in this simulation was lower than those in previous tests, $N=180^2$ particles.} and the lower ones are for the GDSPH calculation. Both look similar, although the axisymmetric calculation is a little less evolved. At $t=1.1\cdot 10^{12}$~s the cloud has collapsed into a disk with similar central density, $\rho\simeq 10^{-12}$\dens, in both cases. Both codes indicate the same qualitative trend with decreasing values of the $\mu$ parameter. A high value, $\mu\rightarrow\infty$~(i.e. $B^z\simeq 0)$ there is no jet at all, whereas the incipient jet is more developed for $\mu=10$, which is encouraging. Following the evolution of the cloud and the jet at longer times is out of the scope of the present work.    

\section{Conclusion}

In this work, we present a novel SPH formulation of ideal magneto-hydrodynamics with axial geometry. The main goal is to tackle problems with higher resolution and lower computational effort than standard SPHMHD codes. The proposed scheme and concomitant hydrodynamic code, called Axis-SPHYNX, have been verified by direct comparison with the results of the three-dimensional SPHMHD code GDSPH, by \cite{wissing20}. On the whole, there is a good match between both hydrodynamic codes in the performed tests. The agreement is excellent in the case of simulating explosions and implosions in magnetized systems, which could be of interest to understanding the physics of plasma compression in terrestrial laboratories. The axisymmetric code is also able to simulate the growth of instabilities such as the Kelvin-Helmholtz instability, which involves longer time-scales than explosions. 

Axis-SPHYNX can handle more complex scenarios such as those involving gravity and rotation of indisputable interest to astrophysics. As shown in Sect.~\ref{subsec:collapse}, with the collapse of a magnetized cloud, the proposed scheme is able to successfully cope with that scenario. Nevertheless, the agreement between both codes is here basically qualitative and work has to be done to enhance the calculations. Immediate prospects are to incorporate  grad-$h$ effects, AV switches, as well as to improve the initial model generation and to refine the treatment of particles that move close to the singularity axis. 

\section*{Acknowledgment}
This work has been supported by the MINECO Spanish project PID2020-117252GB-100, and by the Swiss Platform for Advanced Scientific Computing (PASC) project SPH-EXA: Optimizing Smooth Particle Hydrodynamics for Exascale Computing.  The authors acknowledge the support of sciCORE (http://scicore.unibas.ch/) scientific computing core facility at University of Basel. The GDSPH simulations were performed using the resources from the National Infrastructure for High Performance Computing and Data Storage in Norway, UNINETT Sigma2, allocated to Project NN9477K. We also acknowledge the support from the Research Council of Norway through NFR Young Research Talents Grant 276043.



%

\end{document}